\documentstyle[prc,aps,preprint]{revtex}
\begin{document}
\draft
\date{\today}
\title{Interference effects in the decay $\phi\rightarrow
\pi^{0}\pi^{0}\gamma$ and the coupling constant
g$_{\phi\sigma\gamma}$}

\author{A. Gokalp~\thanks{agokalp@metu.edu.tr} and
        O. Yilmaz~\thanks{oyilmaz@metu.edu.tr}}
\address{ {\it Physics Department, Middle East Technical University,
06531 Ankara, Turkey}}
\maketitle

\begin{abstract}
We study the radiative decay $\phi\rightarrow\pi^{0}\pi^{0}\gamma$
within the framework of a phenomenological approach in which the
contributions of $\sigma$-meson, $\rho$-meson and $f_0$-meson are
considered. We analyze the interference effects between different
contributions and utilizing the experimental branching ratio for
this decay we estimate the coupling constant
$g_{\phi\sigma\gamma}$.
\end{abstract}

\thispagestyle{empty}
~~~~\\
\pacs{PACS numbers: 12.20.Ds, 13.40.Hq }
\newpage
\setcounter{page}{1}
\section{Introduction}
The radiative decays of $\phi$ mesons are valuable sources of
information in low-energy hadron physics in areas such as quark
model, SU(3) symmetry and the Okubo-Zweig-Iizuka (OZI) rule. In
particular, radiative $\phi$ meson decays
$\phi\rightarrow\pi^{0}\pi^{0}\gamma$ and
$\phi\rightarrow\pi^{0}\eta\gamma$ can provide insight on the
structure and properties of low-mass scalar resonances, since
these decays primarily proceed through processes involving scalar
resonances such as $\phi\rightarrow f_0(980)\gamma$ and
$\phi\rightarrow a_0(980)\gamma$, with the subsequent decays into
$\pi^{0}\pi^{0}\gamma$ and $\pi^{0}\eta\gamma$ \cite{R1,R2}. On
the experimental side, the Novosibirsk SND \cite{R3} and CMD-2
\cite{R4} collaborations recently have reported very accurate
results on these decays with the following branching ratios
BR($\phi\rightarrow\pi^{0}\pi^{0}\gamma)=(1.221\pm 0.098\pm
0.061)\times 10^{-4}$,
BR($\phi\rightarrow\pi^{0}\eta\gamma)=(0.88\pm 0.14\pm 0.09)\times
10^{-4}$ \cite{R4}, and
BR($\phi\rightarrow\pi^{0}\pi^{0}\gamma)=(0.92\pm 0.08\pm
0.06)\times 10^{-4}$,
BR($\phi\rightarrow\pi^{0}\eta\gamma)=(0.9\pm 0.24\pm 0.1)\times
10^{-4}$ \cite{R4}, where the first error is statistical and the
second one is systematic.

The low-mass scalar mesons with vacuum quantum numbers have
fundamental importance in understanding low energy QCD
phenomenology and the symmetry breaking mechanisms in QCD. In
addition to $f_0(980)$ and $a_0(980)$, the existence of the
scalar-isoscalar  $\sigma$ meson as a broad $\pi\pi$ resonance,
which has long been controversial, seems to be established.
Increasing number of theoretical and experimental analyzes find a
$\sigma$-pole position near (500-i250) MeV \cite{R5,R6}. A direct
experimental evidence seems to emerge from
$D^+\rightarrow\sigma\pi^{+}\rightarrow 3\pi$ decay channel
observed by the Fermilab (E791) collaboration, where $\sigma$
meson is seen as a clear dominant peak with $M_\sigma$=478 MeV and
$\Gamma_\sigma$=324 MeV \cite{R7}. On the other hand, the nature
and the quark substructure of these scalar mesons have not been
established yet, whether  they are conventional $q\bar{q}$ states
\cite{R8}, $\pi\bar{\pi}$ in case of $\sigma$ \cite{R9} and
$K\bar{K}$ in case of $f_0$ and $a_0$ \cite{R10} molecules, or
exotic multiquark $q^2\bar{q}^2$ states \cite{R11,R12} have been a
subject of debate. It has been noted that the radiative decay of
$\phi$ meson to the scalar mesons can differentiate among various
models of their structure \cite{R1}.

The radiative decay process $\phi\rightarrow\pi^{0}\pi^{0}\gamma$
among other radiative decay processes of the type $V^0\rightarrow
P^{0}P^{0}\gamma$ where V and P belong to the lowest multiplets of
vector (V) and pseudoscalar (P) mesons was studied by Fajfer and
Oakes \cite{R13} using a low energy effective Lagrangian approach
with gauged Wess-Zumino terms. They considered the contributions
in which the virtual vector meson states dominate, and obtained
the branching ratio for this decay as
$BR(\phi\rightarrow\pi^{0}\pi^{0}\gamma)=3.46\times 10^{-5}$. The
contribution of intermediate vector mesons to the decays
$V^0\rightarrow P^{0}P^{0}\gamma$ was later considered by Bramon
et al. \cite{R14} using standard Lagrangians obeying
SU(3)-symmetry, and they obtained the result
$BR(\phi\rightarrow\pi^{0}\pi^{0}\gamma)=1.2\times 10^{-5}$ for
the branching ratio of the $\phi\rightarrow\pi^{0}\pi^{0}\gamma$
decay. Bramon et al. \cite{R15} later considered these decays
within the framework of chiral effective Lagrangians, and using
chiral perturbation theory they calculated the branching ratios
for various decays of the type $V^0\rightarrow P^{0}P^{0}\gamma$
at the one-loop level. They showed that  the one-loop
contributions are finite and to this order no counterterms  are
required. In this approach the decay
$\phi\rightarrow\pi^{0}\pi^{0}\gamma$ proceeds through
charged-kaon loop and they obtained the contribution of
charged-kaon loops to this decay as
$\Gamma(\phi\rightarrow\pi^{0}\pi^{0}\gamma)$=224 eV which is much
larger than the intermediate vector meson state (VMD) contribution
due to OZI rule. Considering the amplitudes of both VMD and
kaon-loop contributions they obtained for the decay rate the value
$\Gamma(\phi\rightarrow\pi^{0}\pi^{0}\gamma)$=269 eV, moreover
they noted that OZI allowed kaon-loops dominate both the photonic
spectrum and the decay rate. Radiative decays of $\phi$ meson were
also investigated by Marco et al. \cite{R16} employing the
techniques of chiral unitary  theory developed earlier by Oller
\cite{R17}. Using a chiral unitary approach, they included the
final state interactions of two pions by summing the kaon-loops
through Bethe-Salpeter equation. They obtained the branching ratio
for the decay $\phi\rightarrow\pi^{0}\pi^{0}\gamma$ as
$BR(\phi\rightarrow\pi^{0}\pi^{0}\gamma)=0.8\times 10^{-4}$.
Moreover, they obtained the photon distribution as a function of
the invariant mass of the $\pi^0\pi^0$ system and compared it with
SND data. They noted that the shape of the experimental spectrum
is relatively well reproduced with the $\phi\rightarrow
f_{0}\gamma$ contribution since $f_0$ meson is the important
scalar resonance appearing in $K^+K^-\rightarrow\pi^0\pi^0$
amplitude. However, they also noted an appreciable strength for a
possible  $\phi\rightarrow\sigma\gamma$ contribution in the
spectrum.

In this work, we follow a phenomenological approach and study the
radiative $\phi\rightarrow\pi^{0}\pi^{0}\gamma$ decay by
considering $\rho$-pole vector meson dominance amplitude as well
as scalar $\sigma$-pole and $f_0$-pole amplitudes. By employing
the experimental value for this decay rate, we estimate the
coupling constant $g_{\phi\sigma\gamma}$. This coupling constant
is an important physical input for studies of $\phi$-meson
photoproduction experiments on nucleons near threshold \cite{R18}.

\section{Formalism}
In our analysis of the radiative decay
$\phi\rightarrow\pi^{0}\pi^{0}\gamma$, we proceed within a
phenomenological framework and we do not make any assumption about
the structure of $f_0$ meson. We note that the $\phi$ and $f_0$
mesons both couple strongly to $K^+K^-$ system, and therefore in
our phenomenological approach we describe the $\phi KK$- and
$f_0KK$-vertices by the effective Lagrangians
\begin{eqnarray}\label{e1}
{\cal L}^{int.}_{\phi KK}=-g_{\phi KK} \phi^\mu (
   K^{+}\partial_{\mu}K^{-}-K^{-}\partial_{\mu}K^{+})~~,
\end{eqnarray}
and
\begin{eqnarray}\label{e2}
{\cal L}^{int.}_{f_0 KK}=g_{f_0 KK}M_{f_0}~ K^{+}K^{-}f_0~~,
\end{eqnarray}
respectively, which also serve to define the coupling constants
$g_{\phi KK}$ and $g_{f_0 KK}$. The  effective Lagrangian for the
$\phi KK$-vertex is the one that results from the standard chiral
Lagrangians  in lowest order of chiral perturbation theory
\cite{R19}. The decay rate for the $\phi\rightarrow K^{+}K^{-}$
decay resulting from this Lagrangian is
\begin{eqnarray}\label{e3}
\Gamma (\phi\rightarrow K^{+}K^{-})= \frac{g^{2}_{\phi
KK}}{48\pi}M_{\phi} \left [
1-\left(\frac{2M_{K}}{M_{\phi}}\right)^{2}\right ] ^{3/2}~~.
\end{eqnarray}
Utilizing the experimental value for the branching ratio
$BR(\phi\rightarrow K^{+}K^{-})=(0.492\pm0.007)$ for the decay
$\phi\rightarrow K^{+}K^{-}$ \cite{R20}, we determine the coupling
constant $g_{\phi KK}$ as $g_{\phi KK}=(4.59\pm 0.05)$.
Furthermore, as a result of the strong coupling of both $\phi$ and
$f_0$ to $K^+K^-$, independent of the nature and dynamical
structure of $f_0$, there is an amplitude for the decay
$\phi\rightarrow f_0\gamma$ to proceed through the charged
kaon-loop which we show diagrammatically in Fig. 1, where the last
diagram assures gauge invariance \cite{R1,R21}. The amplitude of
the radiative decay $\phi\rightarrow f_0\gamma$ that follows from
the diagrams in Fig. 1 is
\begin{eqnarray}\label{e4}
{\cal M}(\phi\rightarrow f_0\gamma)=u^\mu\epsilon^\nu\left (p_\nu
k_\mu-g_{\mu\nu}p\cdot k\right ) \frac{e~g_{\phi
KK}~(g_{f_0KK}M_{f_0})}{2\pi^{2}M_{K}^{2}}~I(a,b)
\end{eqnarray}
where $(u,p)$ and $(\epsilon, k)$ are the polarizations and
four-momenta of the $\phi$ meson and the photon respectively, and
$a=M_{\phi}^{2}/M_{K}^{2}$, $b=M_{f_0}^{2}/M_{K}^{2}$. The
function I(a,b) has been calculated in different contexts
\cite{R2,R17,R22}, and it is given as
\begin{eqnarray}\label{e5}
I(a,b)=\frac{1}{2(a-b)} -\frac{2}{(a-b)^{2}}\left [
f\left(\frac{1}{b}\right)-f\left(\frac{1}{a}\right)\right ]
+\frac{a}{(a-b)^{2}}\left [
g\left(\frac{1}{b}\right)-g\left(\frac{1}{a}\right)\right ]
\end{eqnarray}
where
\begin{eqnarray}\label{e6}
&&f(x)=\left \{ \begin{array}{rr}
           -\left [ \arcsin (\frac{1}{2\sqrt{x}})\right ]^{2}~,& ~~x>\frac{1}{4} \\
\frac{1}{4}\left [ \ln (\frac{\eta_{+}}{\eta_{-}})-i\pi\right
]^{2}~, & ~~x<\frac{1}{4}
            \end{array} \right.
\nonumber \\ && \nonumber \\ &&g(x)=\left \{ \begin{array}{rr}
        (4x-1)^{\frac{1}{2}} \arcsin(\frac{1}{2\sqrt{x}})~, & ~~ x>\frac{1}{4} \\
 \frac{1}{2}(1-4x)^{\frac{1}{2}}\left [\ln (\frac{\eta_{+}}{\eta_{-}})-i\pi \right ]~, & ~~ x<\frac{1}{4}
            \end{array} \right.
\nonumber \\ && \nonumber \\ &&\eta_{\pm}=\frac{1}{2x}\left [
1\pm(1-4x)^{\frac{1}{2}}\right ] ~.
\end{eqnarray}

The decay width for the radiative decay $\phi\rightarrow
f_0\gamma$ can then be obtained from the amplitude ${\cal
M}(\phi\rightarrow f_0\gamma)$ as
\begin{eqnarray}\label{e7}
\Gamma(\phi\rightarrow
f_0\gamma)=\frac{\alpha}{6(2\pi)^4}\frac{M_\phi^2-M_{f_0}^2}{M_\phi^3}
g_{\phi KK}^2~(g_{f_0KK }M_{f_0})^2\mid (a-b)I(a,b)\mid^2
\end{eqnarray}
from which by using the experimental value $BR(\phi\rightarrow
f_0\gamma)=(3.4\pm0.4)\times 10^{-4}$ \cite{R20} we obtain the
coupling constant $g_{f_0 KK}$ as $g_{f_0 KK}=(4.13\pm 1.42)$.

In our phenomenological approach, we assume that the radiative
decay $\phi\rightarrow\pi^{0}\pi^{0}\gamma$ proceeds through the
reactions
$\phi\rightarrow\rho^0\pi^0\rightarrow\pi^{0}\pi^{0}\gamma$,
$\phi\rightarrow\sigma\gamma\rightarrow\pi^{0}\pi^{0}\gamma$, and
$\phi\rightarrow f_0\gamma\rightarrow\pi^{0}\pi^{0}\gamma$. We
write the total amplitude as the sum of the amplitudes of each
reaction and this way we take the interference between different
reactions into account. In order to describe the reaction
$\phi\rightarrow f_0\gamma\rightarrow\pi^{0}\pi^{0}\gamma$ we
again note that both $\phi$ and $f_0$ couple strongly to $K^+K^-$,
furthermore $f_0$ also couples strongly to $\pi^0\pi^0$.
Therefore, we assume that this reaction proceeds by a two-step
mechanism with $f_0$ coupling to $\phi$ with intermediate
$K\bar{K}$ states. We depict the processes contributing to the
$\phi\rightarrow\pi^{0}\pi^{0}\gamma$ decay amplitude
diagrammatically in Fig. 2, where the last diagram in Fig. 2(c)
results from the minimal coupling for gauge invariance.

We describe the $\phi\sigma\gamma$-vertex by the effective
Lagrangian
\begin{eqnarray}\label{e8}
{\cal
L}^{int.}_{\phi\sigma\gamma}=\frac{e}{M_{\phi}}g_{\phi\sigma\gamma}
   \left[\partial^{\alpha}\phi^{\beta}\partial_{\alpha}A_{\beta}
   -\partial^{\alpha}\phi^{\beta}\partial_{\beta}A_{\alpha}\right]\sigma~~,
\end{eqnarray}
which also defines the coupling constant $g_{\phi\sigma\gamma}$.
For the $\sigma\pi\pi$-vertex we use the effective Lagrangian
\begin{eqnarray}\label{e9}
{\cal L}^{int}_{\sigma\pi\pi}=
\frac{1}{2}g_{\sigma\pi\pi}M_{\sigma}\vec{\pi}\cdot\vec{\pi}\sigma~~.
\end{eqnarray}
The decay width of the $\sigma$-meson that follows from this
Lagrangian is given as
\begin{eqnarray}\label{e10}
\Gamma_{\sigma}\equiv\Gamma(\sigma\rightarrow\pi\pi)=
\frac{g^{2}_{\sigma\pi\pi}}{4\pi}\frac{3M_{\sigma}}{8} \left [
1-\left(\frac{2M_{\pi}}{M_{\sigma}}\right)^{2}\right ] ^{1/2}~~.
\end{eqnarray}
In our calculation, for given values of $M_{\sigma}$ and
$\Gamma_\sigma$ we determine the coupling constant
$g_{\sigma\pi\pi}$ by using the expression for $\Gamma_\sigma$.
The $\phi\rho\pi$-vertex in Fig. 2(b) is conventionally described
by the effective Lagrangian
\begin{eqnarray}\label{e11}
{\cal L}^{int.}_{\phi\rho\pi}=g_{\phi\rho\pi}
\epsilon^{\mu\nu\alpha\beta}\partial_{\mu}\phi_{\nu}
\partial_{\alpha}\rho_{\beta}\pi~~.
\end{eqnarray}
The coupling constant $g_{\phi\rho\pi}$ is determined  by N. N.
Achasov and V. V. Gubin \cite{R23} as  $g_{\phi\rho\pi}=(0.811\pm
0.081)~~GeV^{-1}$ using the data on the decay
$\phi\rightarrow\rho\pi\rightarrow\pi^{+}\pi^{-}\pi^0$ \cite{R20}.
The $\rho\pi\gamma$-vertex in Fig. 2(b) is described by the
effective Lagrangian
\begin{eqnarray}\label{e12}
{\cal
L}^{int.}_{\rho\pi\gamma}=\frac{e}{M_{\rho}}g_{\rho\pi\gamma}
\epsilon^{\mu\nu\alpha\beta}\partial_{\mu}\rho_{\nu}
\partial_{\alpha}A_{\beta}\pi~~.
\end{eqnarray}
The coupling constant $g_{\rho\pi\gamma}$ is then obtained from
the experimental partial width of the radiative decay
$\rho\rightarrow\pi^{0}\gamma$ \cite{R20} as
$g_{\rho\pi\gamma}=0.69\pm 0.35$. Finally, we describe the
$f_0\pi^0\pi^0$-vertex by the effective Lagrangian
\begin{eqnarray}\label{e13}
{\cal L}^{int.}_{f_0\pi\pi}=\frac{1}{2}g_{f_0\pi\pi}M_{f_0}
\vec{\pi}\cdot\vec{\pi}f_0~~.
\end{eqnarray}
and the decay width $\Gamma_f$ for the decay
$f_0\rightarrow\pi\pi$ that results from this effective Lagrangian
is given by a similar expression as for $\Gamma_\sigma$. For a
given value of $\Gamma_f$ we use this expression to determine the
coupling constant $g_{f_0\pi\pi}$. Furthermore, in our calculation
of invariant amplitudes we make the replacement $M\rightarrow
M-\frac{1}{2}i\Gamma$ in $f_0$-, $\sigma$-, and $\rho$-propagators
and use the experimental values $\Gamma_{\rho}=(150.2\pm0.8)$ MeV
\cite{R20} for $\rho$-meson, and $\Gamma_\sigma=(324\pm 21)$ MeV
\cite{R7} for $\sigma$-meson. However, the mass $M_{f_0}$=980 MeV
of $f_0$-meson is very close to $K^+K^-$ threshold, and this
induces a strong energy dependence on the width $\Gamma_{f_0}$ of
$f_0$-meson. In order to take that into account different
expressions for $\Gamma_{f_0}$ can be used. A first possibility is
to use an energy dependent width for $f_0$
\begin{equation}\label{e14}
  \Gamma_{f_0}(q^2)=\Gamma^{f_0}_{\pi\pi}(q^2)~\theta(\sqrt{q^2}-2M_\pi)+
\Gamma^{f_0}_{K\overline{K}}(q^2)~\theta(\sqrt{q^2}-2M_K)
 \end{equation}
where $q^2$ is the four-momentum square of the virtual
$f_0$-meson. In this expression the width
$\Gamma^{f_0}_{\pi\pi}(q^2)$ is given as
\begin{equation}\label{e15}
  \Gamma^{f_0}_{\pi\pi}(q^2)=\Gamma^{f_0}_{\pi\pi}~\frac{M_{f_0}^2}{q^2}
  ~\sqrt{\frac{q^2-4M_\pi^2}{M_{f_0}^2-4M_\pi^2}}
\end{equation}
and $\Gamma^{f_0}_{K\overline{K}}(q^2)$ with a similar expression.
We use the experimental value $\Gamma^{f_0}_{\pi\pi}$=40-100 MeV
\cite{R20} and we calculate $\Gamma^{f_0}_{K\overline{K}}$ from
the effective Lagrangian given in Eq. (2). Another and widely
accepted possibility is known from the work of Flatt\'e
\cite{R24}. This amounts to extending the expression for
$\Gamma^{f_0}_{K\overline{K}}(q^2)$ below the $K\overline{K}$
threshold where $\sqrt{q^2-4M_K^2}$ is replaced by
$i\sqrt{4M_K^2-q^2}$ and thus $\Gamma^{f_0}_{K\overline{K}}(q^2)$
becomes purely imaginary. In our work, we consider both
possibilities.

In terms of the invariant amplitude ${\cal M}$(E$_{\gamma}$,
E$_{1}$)= ${\cal M}_a+ {\cal M}_b+ {\cal M}_c$ where ${\cal M}_a$,
${\cal M}_b$, and  ${\cal M}_c$ are the invariant amplitudes
resulting from the diagrams a, b, and c in Fig. 2 respectively,
the differential decay probability for
$\phi\rightarrow\pi^{0}\pi^{0}\gamma$ for an unpolarized
$\phi$-meson at rest is then given as
\begin{eqnarray}\label{e16}
\frac{d\Gamma}{dE_{\gamma}dE_{1}}=\frac{1}{(2\pi)^{3}}~\frac{1}{8M_{\phi}}~
\mid {\cal M}\mid^{2} ,
\end{eqnarray}
where E$_{\gamma}$ and E$_{1}$ are the photon and pion energies
respectively. We perform an average over the spin states of
$\phi$-meson and a sum over the polarization states of the photon.
The decay width $\Gamma(\phi\rightarrow\pi^{0}\pi^{0}\gamma)$ is
then obtained by integration
\begin{eqnarray}\label{e17}
\Gamma=\frac{1}{2}\int_{E_{\gamma,min.}}^{E_{\gamma,max.}}dE_{\gamma}
       \int_{E_{1,min.}}^{E_{1,max.}}dE_{1}\frac{d\Gamma}{dE_{\gamma}dE_{1}}
\end{eqnarray}
where now the factor ($\frac{1}{2}$) is included because of the
$\pi^{0}\pi^{0}$ in the final state. The minimum photon energy is
E$_{\gamma, min.}=0$ and the maximum photon energy is given as
$E_{\gamma,max.}=(M_{\phi}^{2}-4M_{\pi}^{2})/2M_{\phi}$=474 MeV.
The maximum and minimum values for pion energy E$_{1}$ are given
by
\begin{eqnarray}\label{e18}
\frac{1}{2(2E_{\gamma}M_{\phi}-M_{\phi}^{2})} \left\{
-2E_{\gamma}^{2}M_{\phi}+3E_{\gamma}M_{\phi}^{2}-M_{\phi}^{3}
 ~~~~~~~~~~~~~~~~~~~~~~~~~~~~\right. \nonumber \\
\pm  E_{\gamma}\sqrt{(-2E_{\gamma}M_{\phi}+M_{\phi}^{2})
       (-2E_{\gamma}M_{\phi}+M_{\phi}^{2}-4M_{\pi}^{2})}~\left\}\right. ~. \nonumber
\end{eqnarray}

\section{Results and Discussion}
The experimental full width of $f_0$ is $\Gamma_{f_0}$=40-100 MeV
\cite{R20}. Since the coupling constant $g_{\phi\sigma\gamma}$
depends on the value of the width of $f_0$, in order to estimate
its effect on the coupling constant we take this width as
$\Gamma_{f_0}=(70\pm 30)$ MeV and $M_{f_0}$=980 MeV  in our
calculation. If we use the form for
$\Gamma^{f_0}_{K\overline{K}}(q^2)$ that was proposed by Flatt\'e
\cite{R24}, we are not able to reproduce the form of the invariant
mass spectrum for the $\phi\rightarrow\pi^{0}\pi^{0}\gamma$ decay.
In this case the enhancement in the invariant mass spectrum
appears in the central part rather than around the $f_0$ pole. A
similar problem was also encountered by Bramon et al. \cite{R25}
in their study of the role of $a_0(980)$ exchange in the
$\phi\rightarrow\pi^{0}\eta\gamma$ decay. Therefore, in the
analysis presented below, for $\Gamma_{f_0}(q^2)$ we use the form
given in Eq. (14). Through the decay rate that results from the
Lagrangian given in Eq. (13) describing $f_0\pi\pi$-vertex we
obtain the coupling constant $g_{f_0\pi\pi}$ as
$g_{f_0\pi\pi}$=1.58. In order to determine the coupling constant
$g_{\phi\sigma\gamma}$, we follow a similar procedure that we used
in our previous works where we estimated the coupling constants
$g_{\rho\sigma\gamma}$ \cite{R26} and $g_{\omega\sigma\gamma}$
\cite{R27}. We use the experimental value of the branching ratio
for the radiative decay $\phi\rightarrow\pi^{0}\pi^{0}\gamma$
\cite{R3} in our calculation for this decay rate, and this way we
arrive at a quadric equation for the coupling constant
$g_{\phi\sigma\gamma}$, the coefficient of the quadratic term
resulting from $\sigma$ meson contribution shown in Fig. 2(a), and
the coefficient of the linear term from the interference of
$\sigma$-meson amplitude with $\rho$-meson and $f_0$-meson
amplitudes shown in Figs. 2(b) and (c) respectively. Therefore our
analysis produces for a set of values of $\sigma$-meson parameters
$M_\sigma$ and $\Gamma_\sigma$ two values for the coupling
constant $g_{\phi\sigma\gamma}$. We choose for the $\sigma$-meson
parameters the values $M_\sigma=(478\pm 17)$ MeV and
$\Gamma_\sigma=(324\pm 21)$ MeV as suggested by the recent
Fermilab (E791) experiment \cite{R7}, resulting for the coupling
constant $g_{\sigma\pi\pi}$ in the value $g_{\sigma\pi\pi}=5.25\pm
0.32$ using the expression for the decay rate
$\Gamma(\sigma\rightarrow\pi\pi)$ given in Eq. (10). This way we
obtain for the coupling constant $g_{\phi\sigma\gamma}$ the values
$g_{\phi\sigma\gamma}=0.064\pm 0.008$ and
$g_{\phi\sigma\gamma}=0.025\pm 0.009$. We then study the invariant
mass distribution for the reaction
$\phi\rightarrow\pi^{0}\pi^{0}\gamma$. In Fig. 3 we plot the
distribution $dBR/dM_{\pi\pi}=(M_{\pi\pi}/M_{\phi})dBR/dE_\gamma$
for the radiative decay $\phi\rightarrow\pi^{0}\pi^{0}\gamma$ in
our phenomenological approach choosing the coupling constant
$g_{\phi\sigma\gamma}$=0.064 and in Fig. 4 we show the same
distribution for $g_{\phi\sigma\gamma}$=0.025 as a function of the
invariant mass $M_{\pi\pi}$ of $\pi^0\pi^0$ system. In these
figures we also indicate the contributions coming from the
different reactions
$\phi\rightarrow\sigma\gamma\rightarrow\pi^{0}\pi^{0}\gamma$,
$\phi\rightarrow\rho^0\pi^0\rightarrow\pi^{0}\pi^{0}\gamma$, and
$\phi\rightarrow f_0\gamma\rightarrow\pi^{0}\pi^{0}\gamma$ as well
as the contribution of the total amplitude which includes the
interference terms as well. The difference between the two total
contribution curves is mainly in the interference region
$M_{\pi\pi}<0.7$ GeV above which $f_0$-amplitude dominates the
spectrum. From the analysis of the spectrum obtained with the
coupling constants $g_{\phi\sigma\gamma}$=0.064 and
$g_{\phi\sigma\gamma}$=0.025 in Figs. 3 and 4, respectively, we
may decide in favor of the latter value of the coupling constant
$g_{\phi\sigma\gamma}$ and we may state that the experimental data
within the framework of our phenomenological approach suggest that
$g_{\phi\sigma\gamma}=0.025\pm 0.009$. In Fig. 5, we show the
contributions of different amplitudes and the contributions of the
interference terms in interference region $M_{\pi\pi}<0.7$ GeV for
$g_{\phi\sigma\gamma}$=0.025.

On the other hand, the photoproduction of $\rho^0$-meson on proton
targets near threshold can be described at low momentum transfers
by single one-meson exchange model. Friman and Soyeur \cite{R28}
showed that in this picture the  $\rho^0$-meson photoproduction
cross-section on protons is given mainly by $\sigma$-exchange.
They calculated $\rho\sigma\gamma$-vertex assuming vector meson
dominance of the electromagnetic current, and their result when
described using an effective Lagrangian for the
$\rho\sigma\gamma$-vertex gives the value
$g_{\rho\sigma\gamma}\approx$2.71 for this coupling constant.
Later, Titov et al. \cite{R18} in their study of the structure of
$\phi$-meson photoproduction amplitude based on one-meson exchange
and Pomeron-exchange mechanism used the coupling constant
$g_{\phi\sigma\gamma}$ which they calculated from the above value
of $g_{\rho\sigma\gamma}$ invoking unitary symmetry arguments as
$g_{\phi\sigma\gamma}\approx$0.047. Our result for this coupling
constant $g_{\phi\sigma\gamma}=0.025\pm 0.009$ is not in total
agreement with this value. We note that in order to derive  their
result Titov et al. assumed that $\sigma$, $f_0$, and $a_0$ are
members of a unitary nonet. However, assignments of scalar states
into various unitary nanets are not without problems and other
possible assignments than used by Titov et al. have also been
suggested \cite{R29}. In our work we do not make any assumption
about the quark substructure of $\sigma$ and $f_0$ mesons and
describe their couplings in a phenomenological framework.

N. N. Achasov and V. V. Gubin \cite{R23} performed a fit to the
experimental data and they obtained the following values: the
branching ratio with interference $BR(\phi\rightarrow
(f_0\gamma+\pi^0\rho^0)\rightarrow\pi^{0}\pi^{0}\gamma)=1.26\times
10^{-4}$, the branching ratio of the signal $BR(\phi\rightarrow
f_0 \gamma\rightarrow\pi^{0}\pi^{0}\gamma)=1.01\times 10^{-4}$,
the branching ratio of the background
$BR(\phi\rightarrow\rho^0\pi^0\rightarrow\pi^{0}\pi^{0}\gamma)=0.18\times
10^{-4}$. If we use the coupling constant
$g_{\phi\sigma\gamma}$=0.025 and $M_\sigma$=478 MeV
$\Gamma_\sigma$=324 MeV, we obtain for the branching ratios for
different contributing reactions to the radiative decay
$\phi\rightarrow\pi^{0}\pi^{0}\gamma$ the values
$BR(\phi\rightarrow
f_0\gamma\rightarrow\pi^{0}\pi^{0}\gamma)=1.29\times 10^{-4}$,
$BR(\phi\rightarrow\sigma\gamma\rightarrow\pi^{0}\pi^{0}\gamma)=0.04\times
10^{-4}$,
$BR(\phi\rightarrow\rho^0\pi^0\rightarrow\pi^{0}\pi^{0}\gamma)=0.14\times
10^{-4}$, $BR(\phi\rightarrow
(f_0\gamma+\pi^0\rho^0)\rightarrow\pi^{0}\pi^{0}\gamma)=1.34\times
10^{-4}$, $BR(\phi\rightarrow
(f_0\gamma+\sigma\gamma)\rightarrow\pi^{0}\pi^{0}\gamma)=1.16\times
10^{-4}$ and for the total interference term
$BR$(interference)=$-0.25\times 10^{-4}$. Our results are in
reasonable agreement with those obtained in the analysis of N. N.
Achasov and V. V. Gubin \cite{R23}, with the difference that our
results include contributions coming from the $\sigma$-pole
amplitude as well as $\rho^0$- and $f_0$-pole amplitudes.

As we already noted, the spectrum for the decay
$\phi\rightarrow\pi^0\pi^0\gamma$ is dominated by the
$f_0$-amplitude, and the contribution coming from the
$\sigma$-amplitude can only appreciably be noticed in the region
$M_{\pi\pi}<0.7$ GeV through interference effects. Our analysis of
these interference effects and our calculation for the decay rate
of the radiative decay $\phi\rightarrow\pi^0\pi^0\gamma$ suggests
the value of the coupling constant $g_{\phi\sigma\gamma}$ as
$g_{\phi\sigma\gamma}=0.025\pm 0.009$.

\newpage

{\bf Figure Captions:}

\begin{description}
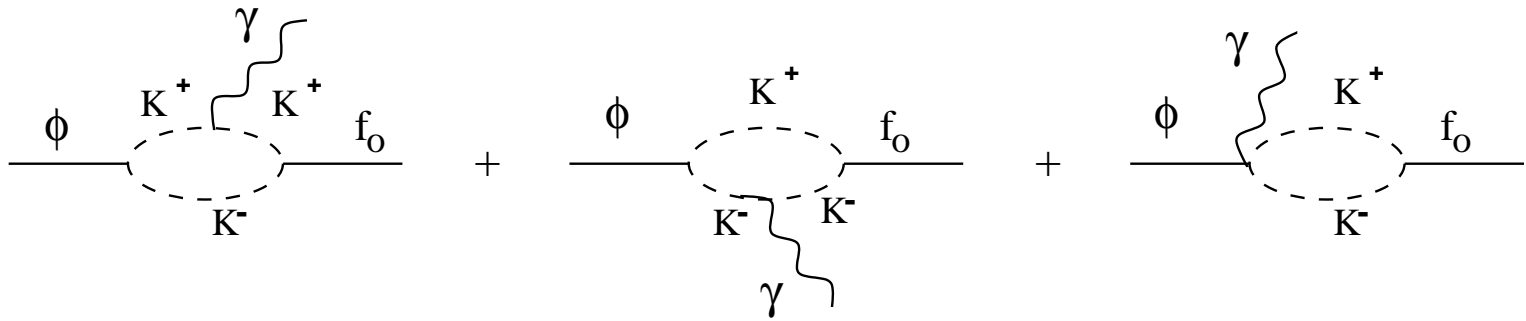
\item[{\bf Figure 1}:] Diagrams for the decay
$\phi\rightarrow f_{0}\gamma$
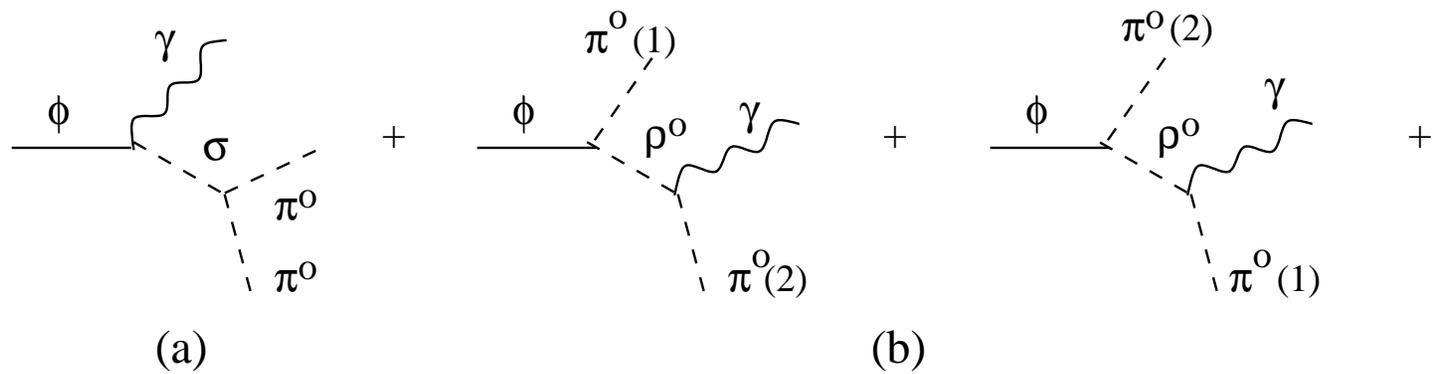
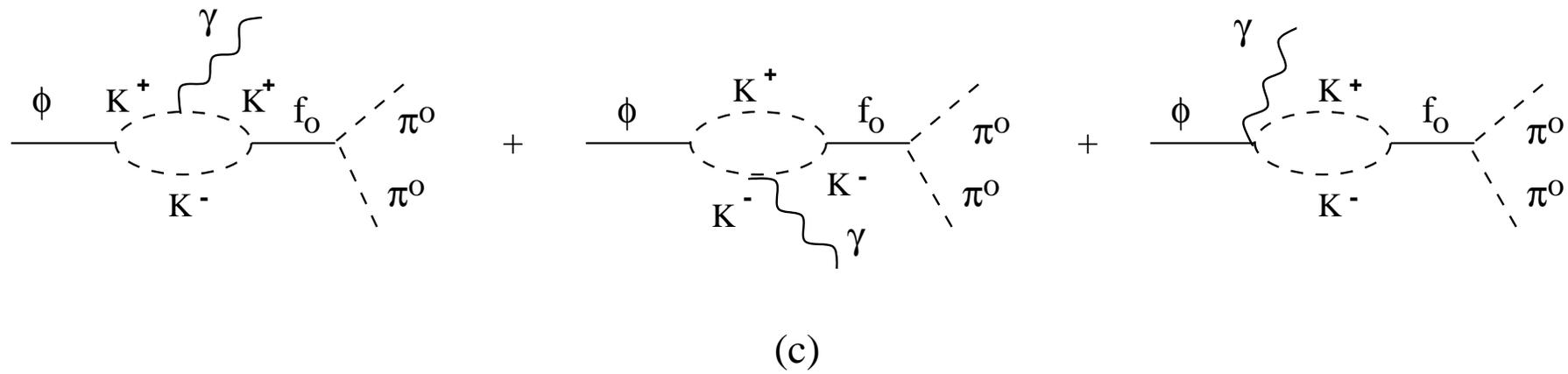
\item[{\bf Figure 2}:] Diagrams for the decay
$\phi\rightarrow \pi^0\pi^0\gamma$
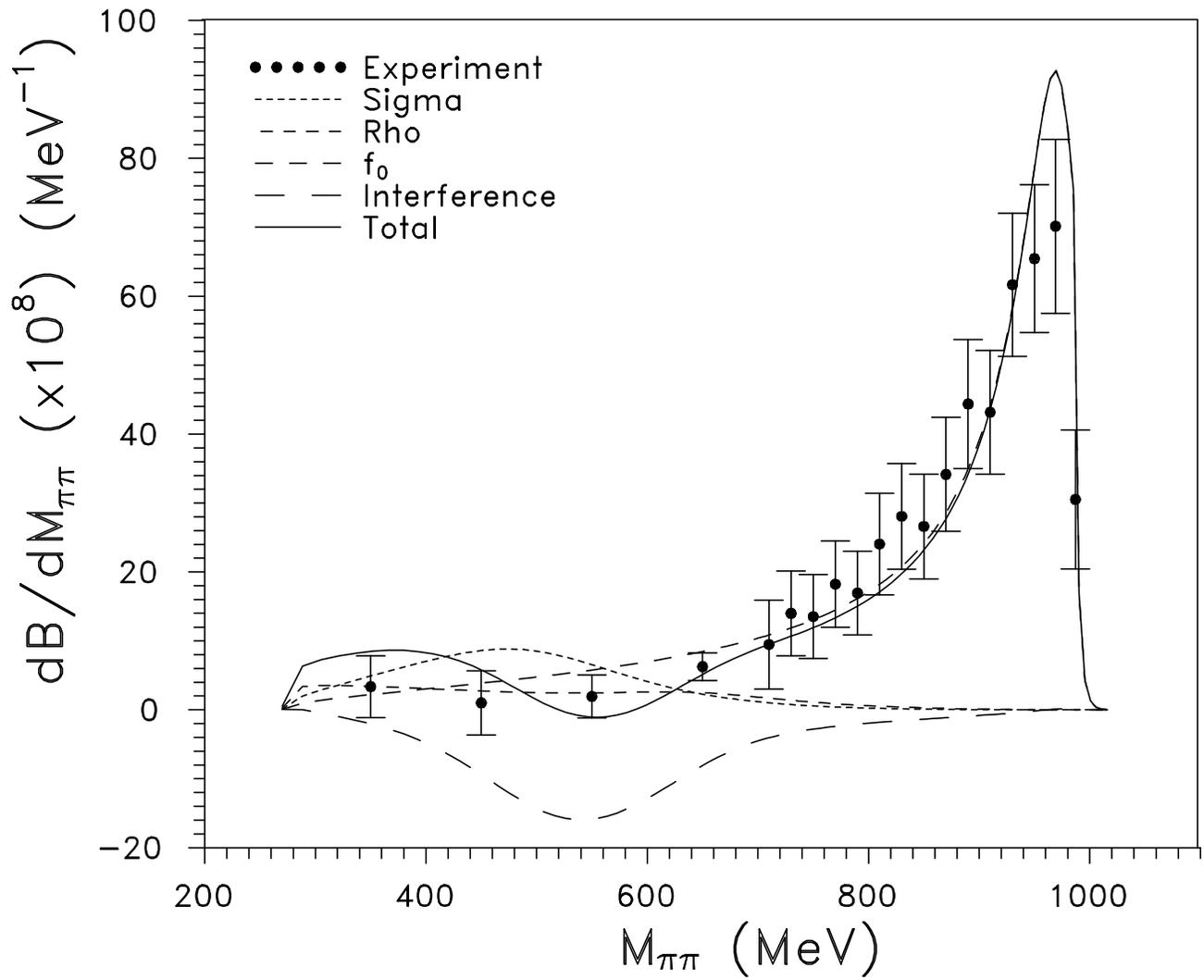
\item[{\bf Figure 3}:] The $\pi^0\pi^0$ invariant mass spectrum for the decay
$\phi\rightarrow\pi^{0}\pi^{0}\gamma$ for
g$_{\phi\sigma\gamma}$=0.064. The contributions of different terms
are indicated.
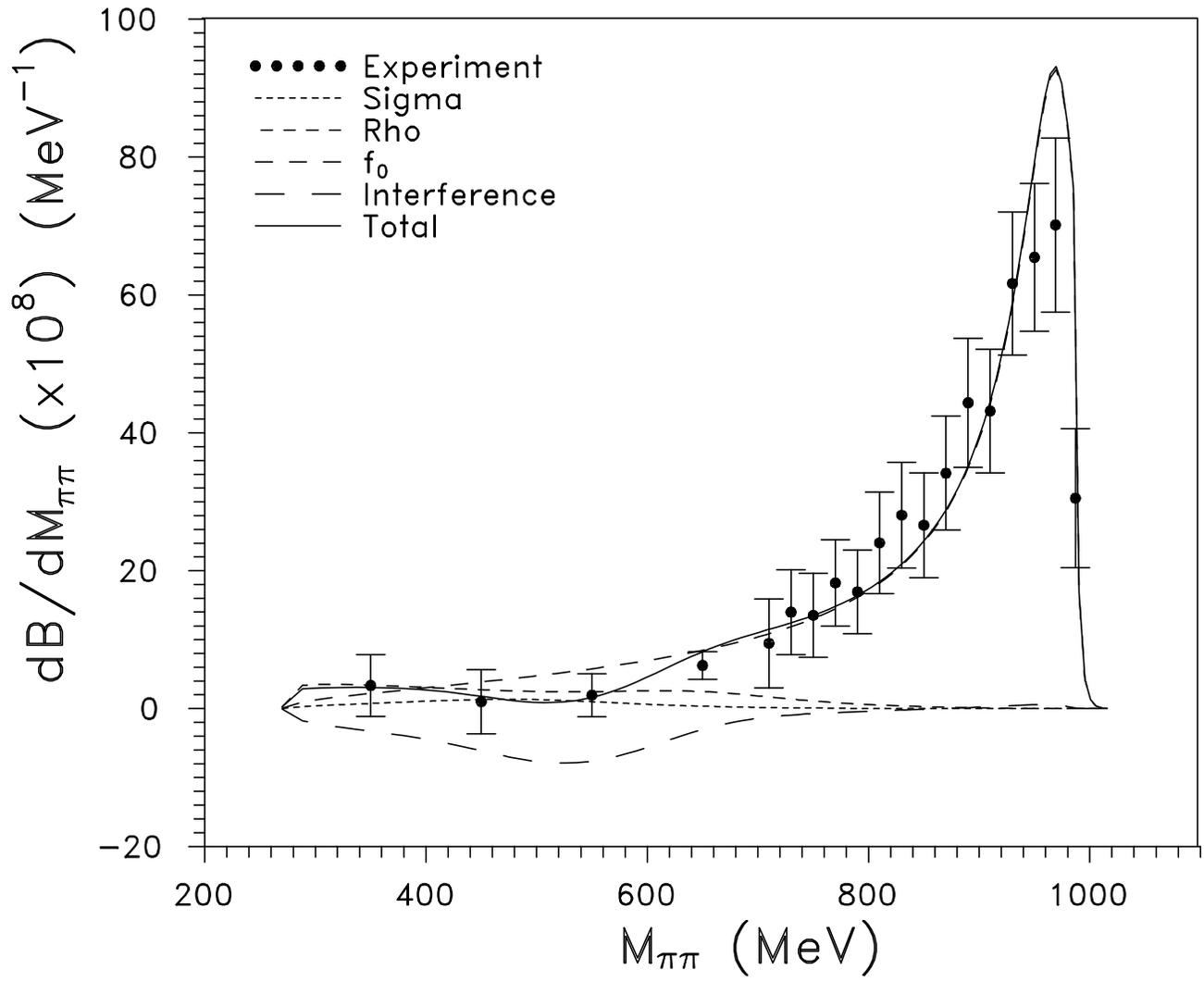
\item[{\bf Figure 4}:] The $\pi^0\pi^0$ invariant mass spectrum for the decay
$\phi\rightarrow\pi^{0}\pi^{0}\gamma$ for
g$_{\phi\sigma\gamma}$=0.025. The contributions of different terms
are indicated.
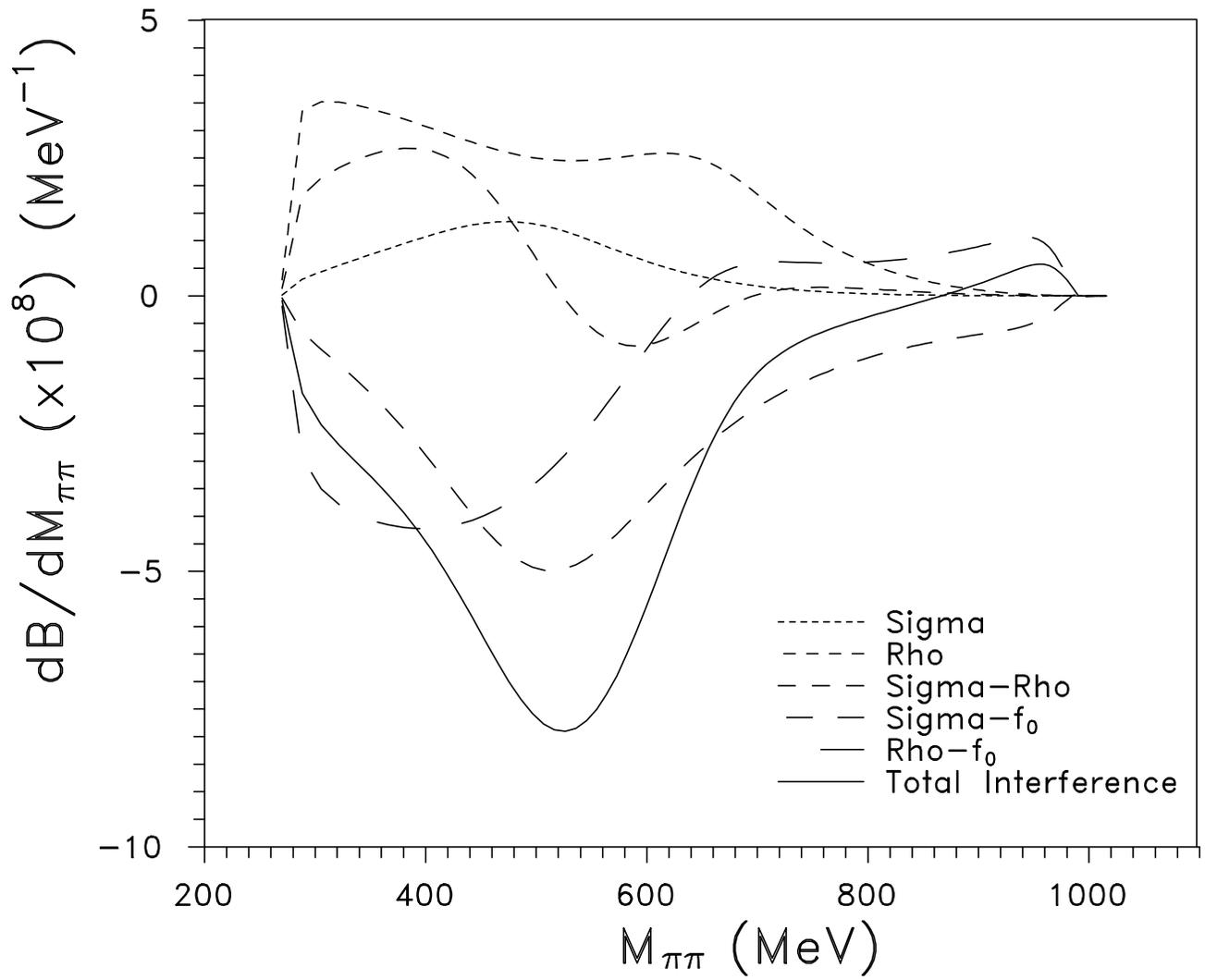
\item[{\bf Figure 5}:] The contributions of $\rho$- and $\sigma$- amplitudes and
 interference terms to the invariant mass spectrum of the decay
$\phi\rightarrow\pi^{0}\pi^{0}\gamma$ for
g$_{\phi\sigma\gamma}$=0.025.
\end{description}

\end{document}